\documentclass[a4paper,final]{appolb}
\usepackage{amsmath}
\usepackage{amssymb}

\usepackage{graphicx}
\begin{document}

\title{CuBA - a CUDA implementation of BAMPS
\thanks{Presented by Ulrike Eilhauer at the International Meeting "Excited QCD", Peniche, Portugal, 6 - 12 May, 2012}}

\author{Nuno Cardoso, Pedro Bicudo, Ulrike Eilhauer
\address{CFTP, Departamento de F\'{i}sica, Instituto Superior T\'{e}cnico, Universidade T\'{e}cnica de Lisboa, Av. Rovisco Pais, 1049-001 Lisbon, Portugal\newline}
\\
{Ioanni Bouras
}
\address{Institut fuer Theoretische Physik, Johann Wolfgang Goethe-Universitaet, Max-von-Laue-Strasse 1, D-60438, Frankfurt am Main, Germany}
}

\maketitle

\begin{abstract}

Using CUDA as programming language, we create a code named CuBA which is based on the CPU code "Boltzmann Approach for Many Parton Scattering (BAMPS)" developed in Frankfurt in order to study a system of many colliding particles resulting from heavy ion collisions.
Furthermore, we benchmark our code with the Riemann Problem and compare the results with BAMPS. They demonstrate an improvement of the computational runtime, by one order of magnitude.

\end{abstract}
\PACS{11.15.Ha; 12.38Gc; 12.38Mh}

\section{Introduction}

Basing ourselves on the BAMPS code developed in Frankfurt by C.Greiner, Z.Xu et al., we decided to study the interaction between the gluons of a gluon gas produced at the onset of Heavy Ion Collisions \cite{Bouras:2009}. We use CUDA as programming language to create the code CuBA "The Boltzmann Approach for Many Parton Scattering written with CUDA" \cite{CUFFT}. We expect to get an improvement of the computational runtime. In addition, both codes are benchmarked with the Riemann problem to compare the results of the two programs. 

In this paper we investigate the physical concepts behind this program, the CUDA language and finally the prior results obtained.

\section{Theory}
We are specially interested in solving the Riemann problem in viscous matter using the relativistic Boltzmann equation which is as follows,

\begin{equation}
	\left( \frac{\delta}{\delta t}+\frac{p_1}{m}\cdot \nabla_r + F \cdot \nabla_{p1}\right)f_1=\int{d^3p_2d^3p_1^{'}d^3p_2^{'}\delta^4 (P_f-P_i)\left|T_{fi}\right|^2 (f_2^{'}f_1^{'}-f_2f_1)}
\end{equation}

To get a good compromise between computational runtime and physical accuracy, we used the application of microscopic theories together with strong assumptions like neglecting quantum mechanical effects. 

The main idea for solving the Boltzmann equation with the Particles-In-A-Cell-method (PIC) consists in dividing a certain volume into many cells with volume $V_{cell}= \Delta x \Delta y \Delta z$, where we have N particles, which will suffer movement- and collision-laws in a certain time interval $\Delta t$. Each particle will have its own position $r$ and momentum $p$. So if the particle does not collide, its propagation is given by,

\begin{equation}
x\mapsto x+v_x\Delta t= x+c^2\frac{p_x}{E}\Delta t	
\end{equation}

The same is valid for the y and z directions.

On the other hand, it is important to consider that the collisions are binary and can only occur between particles in the same cell. Therefore, the probabilty of collisions to occur is given by using the Monte-Carlo method in 
$\Delta t$,

\begin{equation}
P_{22}=v_{rel} \frac{\sigma}{N_{test}} \frac{\Delta t}{V_{cell}}	
\end{equation}

 being $\sigma$ the total cross section, which is considered to be isotropic and $v_{rel}$ the relative velocity given by,
$v_{rel}=\frac{s}{2E_1E_2}$ 
where s is the Mandelstam variable, $s=(p_1+p_2)^2$ \cite{Bouras:2009}. 

To reduce statistical fluctuations and to keep the accuracy of our pretended solution, we use the testparticle method. It consists in introducing $N_{test}=r_{test}N$ with $r_{test}$ as a chosen factor, which increases the number of particles. To keep the mean free path $\lambda$ independent of $N_{test}$ we reduce the probaility $P_{22}$ by the same $r_{test}$.
To get the direction of the outgoing momentum we boost from the plasma frame to the center of mass frame applying the Lorentz transformation. In the center of mass frame we choose the momentum randomly. After that, we boost back to the orginial frame.
If a particle collides with one of the six walls established by the box volume, it will be elastically reflected.

\section{CUDA language}
Cuda is a language for parallel programming in gpus, which recently started being used in numerical computations in physics, due to the potential performance increased by order of magnitude.

The CUDA logic is built by writing kernel functions, which calculate the physical matters, in the device and calling them using the host. The device is constituted by various grids which include about 65535$^3$ blocks for Fermi arquitecures and 65535$^2$ blocks for older arquitectures. Each block has 256 threads. The postion and momentum of each particle in $\Delta t$ is stored in a thread. So we point out that the big advantage of using CUDA consists in the fast shared memory region that can be shared among threads \cite{Cudabook1} \cite{Cudabook2} .

\section{Flowchart}

\begin{figure}[htp]
\vspace{-.5cm}
\begin{center}
\includegraphics[scale=0.4]{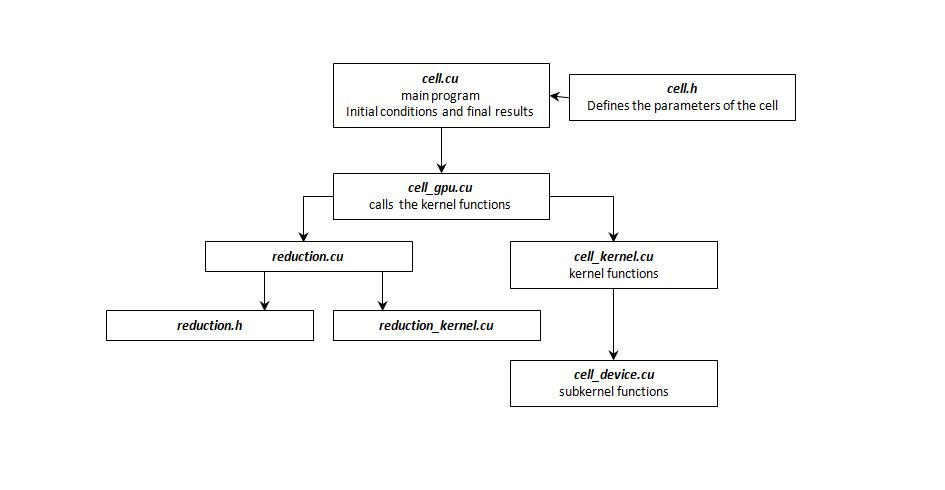}
\end{center}
\vspace{-1.cm}
\caption{Flowchart of our CuBA code.}
\label{Figure 1}
\end{figure}

Our code structure is presented in figure \ref{Figure 1}.

\section{Results}

To test our code we have to take into account the initial conditions we choose. The two important parameters are the time variation $\Delta t$ and x variation $\Delta x$,
once we consider a transverse homogeneous plan. 
$\Delta t$ is always choosen to be smaller than $\Delta x$
to avoid large local variations in one time step.
If we increase $\Delta x$, we have to increase the testparticle number $N_{test}$.
The more testparticles we have, more the curve of the Riemann problem 
approximates to the theoretical solution.
A small testparticle number affects the fluctuations.
To simulate an ideal fluid we may choose a very small viscosity.

\begin{figure}[htp]
\centering
\includegraphics[scale=0.5]{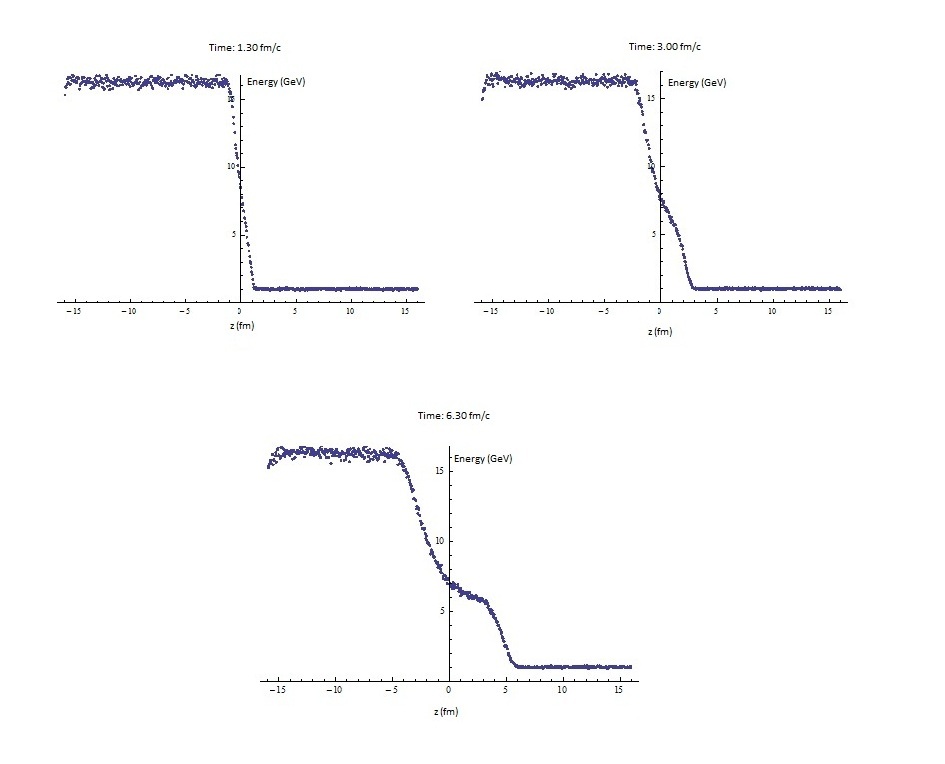}
\caption{Evolution of the energy
density, shown for different time-slices  $\Delta t$. The propagation
of the two waves from the initial boundary of the Riemann problem is
clearly visible.}
\label{Figure 2}
\end{figure}

First, we check some numerical solutions for CuBA considering various parameters. For starters we consider our box volume to be 32$^{3}$ fm$^3$, the cross secion, $\sigma$= 10 GeV$^{-2}$ , dt=0.1 fm/c, with equal particle distribuition at the beginning and diferent temperatures on each side of the box, T$_{left}$=0.4 GeV and T$_{right}$=0.2 GeV. The conservation of the total energy is verified, just as it was expected. We observe the evolution in $\Delta t$ in figure \ref{Figure 2}.

In addition, we observe in figure \ref{Figure 2} the typical figure of the Riemann problem. This problem consists of a propagating shock wave because the initial conditions impose different temperatures \cite{Bouras:2010hm}.

\begin{figure}[htp]
\centering
\includegraphics[scale=0.5]{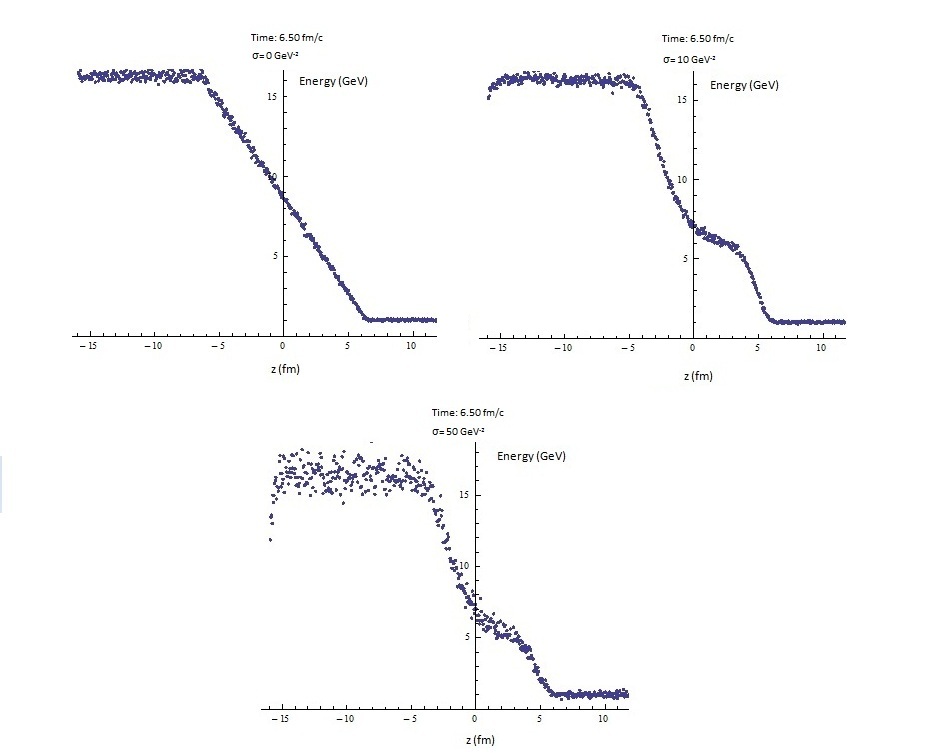}
\caption{Evolution of the local cross
section, shown for different time-slices.}
\label{Figure 3}
\end{figure}

Secondly, we range the cross section, considering the other variables constant and as previously refered. We observe the diferences in figure \ref{Figure 3}.

As we can verify, the slope undoes itself by increasing the cross section, which physically means to have a larger viscosity.

At last, to compare our results to the BAMPS code we choose the same initial conditions in both codes, which are the ones mentioned at the beginning of this section.

\begin{figure}[htp]
\centering
\includegraphics[scale=0.25]{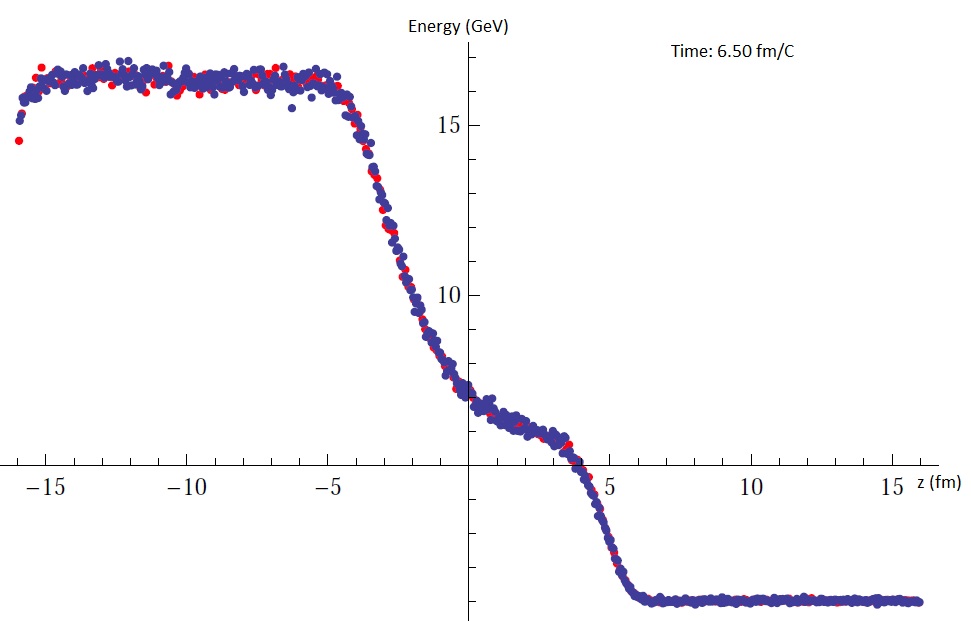}
\caption{Comparing the energy density of
BAMPS  (red points) with CuBA (blue points), both codes produce the
same results, except for statistical fluctuations.}
\label{Figure 4}
\end{figure}

In figure \ref{Figure 4} we can surely identify the overlapping of the results obtained with CuBA (blue points) and BAMPS (red points).

While the BAMPS code spents 12 minutes and 36 seconds to calculate the data, CUBA just needs 58.09 seconds.

\section{Conclusions}

The resulting data can be used to confirm the CPU code and improve the study of shocking particles.
For now we can say that CuBA is about 13 times faster than BAMPS.

In the near future we pretend to implement the parameter $dt$ as variable and optimize our code in computational runtime. Furthermore we will check our code with other initial conditions and compare it to BAMPS. 

As final result we expect to obtain a code which is able to calculate any problem of this type and being as fast as cuda allows us. 

\section*{Acknowledgments}
This work was financed by the FCT contracts POCI/FP/81933/2007, CERN/FP/83582/2008, PTDC/FIS/100968/2008, CERN/FP/109327/2009, NVIDIA Academic Partnership and the CRUP/DAAD exchange A- 10/10.
Nuno Cardoso is also supported by FCT under the contract SFRH/BD/44416/2008.


\end{document}